\documentclass{ifacconf}

	\usepackage{graphics}      
\usepackage{natbib}        
\usepackage{amsmath}
\usepackage{amsfonts}
\usepackage{subcaption}
\usepackage{epsfig}
\usepackage{algorithm}
\usepackage{algpseudocode}
\usepackage{csquotes}
\usepackage{verbatim}
\usepackage{booktabs}
\usepackage{subcaption}
\usepackage{multirow}
\usepackage{mathtools}
\begin{document}
\begin{frontmatter}

\title{Learning the cost-to-go for mixed-integer nonlinear model predictive control}

\author[First,Second]{Christopher A. Orrico}
\author[First]{W.P.M.H. Heemels}
\author[First]{Dinesh Krishnamoorthy}

\address[First]{Eindhoven University of Technology, 5600 MB Eindhoven, The Netherlands (e-mail: c.a.orrico@tue.nl, w.p.m.h.heemels@tue.nl, d.krishnamoorthy@tue.nl).}
\address[Second]{DIFFER - Dutch Institute for Fundamental Energy Research, 5612 AJ Eindhoven, The Netherlands (e-mail: c.orrico@differ.nl)}

\begin{abstract}
	Application of nonlinear model predictive control (NMPC) to problems with hybrid dynamical systems, disjoint constraints, or discrete controls often results in mixed-integer formulations with both continuous and discrete decision variables. However, solving mixed-integer nonlinear programming problems (MINLP) in real-time is challenging, which can be a limiting factor in many applications. To address the computational complexity of solving mixed integer nonlinear model predictive control problem in real-time, this paper proposes an approximate mixed integer NMPC formulation based on value function approximation. Leveraging Bellman's principle of optimality, the key idea here is to divide the prediction horizon into two parts, where the optimal value function of the latter part of the prediction horizon is approximated offline using expert demonstrations. Doing so allows us to solve the MINMPC problem with a considerably shorter prediction horizon online, thereby reducing the online computation cost. The paper uses an inverted pendulum example with discrete controls to illustrate this approach.
\end{abstract}


\end{frontmatter}

Model predictive control has emerged as a powerful tool to control  dynamic systems  in a wide range of application areas, due to its ability to explicitly handle constraints and optimize  desired  performance criterion. This is done by formulating the control problem as a discrete time finite horizon optimal control problem, which is  repeatedly solved online at each sampling instant, given the current state of the system. the first control input from the optimal trajectory is implemented in the system and the process is repeated.

As the domain of MPC application expands to new class of problems, this requires more complex optimization problem formulations. Problem classes that commonly arises in many application domains include, discrete actuators (where the controls only take values from a finite set), switching systems (where the  system dynamics are discontinuous and non-smooth), problems with disjoint constraint sets, avoidance constraints in path planning, or problems with specified logics etc \citep{richards2005MIP-MPC}.
MPC framework can  be applied to such problem classes by formulating the optimal control problem with both continuous and integer decision variables. The resulting nonlinear MPC framework with discrete decision variables require the online solution of mixed integer nonlinear programming (MINLP) problems at each sampling instant. However, MINLP problems in general are $ \mathcal{NP} $-hard, which makes it challenging for real-time control.

Model predictive control, as the name suggests, uses a model to predict the effect of the control actions over a finite rolling horizon, and chooses an optimal control sequence that optimizes a given performance criterion.
Ideally, we want a sufficiently long prediction horizon, since this would help the controller to consider the effect of the control actions far into the future, resulting in a better control policy. However, the size of the optimization problem, and in turn the computational cost, increases with the length of the prediction horizon. This is only further amplified in the presence of discrete decision variables, where the  number of nodes in the branch and bound algorithms increases. Using a short prediction horizon naively on the other hand leads to myopic policies that can result in performance degradation.

This paper argues that in want for very short prediction horizons all is not lost, and that value function approximations enable us to leverage Bellaman's principle of optimality, such that the online controller can be reduced to even as small as a one-step-look-ahead controller without jeopardizing the control performance.
The prediction horizon in the original MINMPC problem can be divided into two parts, where the optimal value function of the tail part of the prediction horizon is replaced by a cost-to-go function. Since the optimal cost-to-go  function is not known exactly, this can be approximated offline, and the online controller is solved with the approximate value function as its cost-to-go. Replacing the optimal value function with approximate cost-to-go functions are also commonly referred to as \textit{approximate dynamic programming} (ADP) \cite{bertsekas2011dynamic}.

This paper uses inverse optimization to learn the value function from expert demonstrations comprising of optimal state-action pairs \citep{keshavarz2011imputing}. The  expert demonstrations may either be from the original long horizon MINMPC solved offline, or other approximate MINMPC strategies, or may even be from human expert demonstrations. Learning the cost function from (near) optimal state-action pairs is broadly studied under the context of learning from demonstrations (LfD). Unlike other learning from demonstration approaches such as direct policy fitting/imitation learning, learning the cost function via inverse optimization requires considerably much lesser data that need not cover the entire feasible state space.
To this end, the main contribution of this paper is a mixed integer NMPC formulation based on value function approximation, where we show that the myopic mixed integer NMPC can be formulated with  very short prediction horizons that results in comparable performance as a long prediction horizon, at a fraction of the online computation cost.

\paragraph*{Related work} 
Various strategies and heuristics have been proposed in the literature to address real-time mixed-integer Nonlinear Model Predictive Control (NMPC) problems. These include solving the relaxed problem by removing the integrality constraint and obtaining the integer trajectory through rounding schemes like Sum-Up-Rounding \citep{sager2009reformulations}. Another approach involves a penalty term homotopy method, where the integer variable is replaced with the constraint $ z(1-z) \le \beta, \; z \in [0,1]  $ along with the homotopy $ \beta \rightarrow 0^+ $ \citep{sager2006numerical}. Nonlinear functions are often approximated using piecewise linearization to transform the Mixed-Integer Nonlinear Programming (MINLP) into Mixed-Integer Linear Programming (MILP) or Mixed-Integer Quadratic Programming (MIQP). However, such approximations may not always be applicable. Real-time iteration schemes for MINMPC based on outer convexification and rounding schemes have been extended \citep{deMauri2020real}, though this becomes challenging if discrete variables appear in inequality constraints. Recent interest in deep learning has led to direct policy approximation approaches \cite{karg2018deep}, where the MPC policy is approximated using deep neural networks trained on large amounts of offline-generated data. However, this requires extensive data covering the entire feasible state space and poses challenges for any online updates and changes in the controller.


\section{Problem formulation}
\label{sec:problemFormulation}
Consider a mixed integer nonlinear MPC problem
\begin{subequations}\label{Eq:MINMPC}
	\begin{align}
		V^*(x(t)) = \min \; & \sum_{k=0}^{N-1}\ell(x_{k},u_{k},z_{k}) + \ell_N(x_{N}) \\
		\text{s.t.} \; & x_{k+1} = f(x_{k},u_{k},z_{k})   \\
		& u_{k} \in \mathcal{U}, \quad z_{k} \in \mathcal{Z}\\
		& x_{0} = x(t)
	\end{align}
\end{subequations}
where $ x \in \mathbb{R}^{n_{x}} $ are the set of states, $ u \in \mathcal{U} \subseteq \mathbb{R}^{n_{u}} $ denotes the set of continuous controls, $ z \in \mathcal{Z} \subseteq \mathbb{Z}^{n_{z}} $ denotes the set of discrete controls, $ N $ denotes the length of the prediction horizon, $ \ell: \mathcal{X} \times \mathcal{U} \times\mathcal{Z}  \rightarrow \mathbb{R}$ and $ \ell_N: \mathcal{X}  \rightarrow \mathbb{R}  $ denote the stage and terminal costs respectively, and  $ f: \mathcal{X} \times \mathcal{U} \times\mathcal{Z}  \rightarrow \mathbb{R}^{n_{x}} $ denotes the system dynamics.
Solving the MINLP \eqref{Eq:MINMPC} at each sampling instant and implementing the first control elements $ [u_{0},z_{0}]^{\mathsf{T}} $ results in the mixed integer MPC policy $
	[u(t),z(t)]^{\mathsf{T}} = 	\pi_{\text{mpc}}(x(t))$.

\textit{Problem setting:} The MINMPC problem \eqref{Eq:MINMPC} is designed to be optimal, but  computationally intensive and is difficult to solve online within the desired sampling time.

\textit{Objective:} Using the MINMPC problem \eqref{Eq:MINMPC} as the \enquote{expert} that can be queried offline, we want to build a simpler control policy that is amenable for online implementation.

The optimal value function of the OCP starting from $ x_{1} $ can be denoted by
\begin{subequations}
	\begin{align}
		V_{\eta}= \min \; & \sum_{k=1}^{N-1}\ell(x_{k},u_{k},z_{k}) + \ell_N(x_{N}) \\
		\text{s.t.} \; & x_{k+1} = f(x_{k},u_{k},z_{k})   \\
		&  g(x_{k},u_{k},z_{k}) \leq 0\\
		&\qquad \forall k = 1,\dots,N-1 \nonumber
	\end{align}
\end{subequations}
That is $ V_{\eta} $ is the optimal-cost-to-go when starting from state $ x_{1}$.
Then by  Bellman's principle of optimality, we can truncate the prediction horizon and  equivalently solve
\begin{subequations}\label{Eq:ADPMINMPC1}
\begin{align}
	\min_{u,z} \; & \ell(x(t),u,z) + V_{\eta}(f(x(t),u,z))  \\
	\text{s.t.} \; & g(x(t),u,z)\leq 0
\end{align}
\end{subequations}
which is significantly easier to solve than the original MINMPC problem \eqref{Eq:MINMPC}. By simply reducing the length of the prediction horizon, and consequently, the number of decision variables, mixed integer problems can be solved more efficiently.

The challenge with the DP recursion is that calculating the optimal cost-to-go $ V_{\eta} $ from all possible states $ x \in \mathcal{X} $ can be intractable and prohibitively time consuming. This can be addressed using a class of methods known as approximate dynamic programming (ADP), where one can approximate the optimal cost-to-go function $ V_{\eta} $ in \eqref{Eq:ADPMINMPC1} by a convex function approximation $ \mathcal{V}(x):= x^{\mathsf{T}} \mathbf{P} x$, parameterized by $ \mathbf{P} $.
The task is then to find $  \mathbf{P} $ such that the myopic policy obtained by solving \eqref{Eq:ADPMINMPC1} mimics the full horizon MINMPC policy \eqref{Eq:MINMPC}.
To do this, we first generate a set of optimal state-action pairs $ \mathcal{D}:= \{(x_{i},[u_{i},z_{i}]^{\mathsf{T}})\}_{i=1}^M $ by solving the full horizon MINMPC offline. This dataset is then used to \textit{impute} the parameter $ \mathbf{P}  $ of the myopic policy, such that the dataset $ \mathcal{D} $ is approximately consistent with the data set $ \mathcal{D} $. To do this, we use inverse optimization.
\paragraph*{Imputing the cost-to-go function:}
Denoting the set of  decision variables of the myopic MPC \eqref{Eq:ADPMINMPC1} by $ \mathbf{w}:= [u,z]^{\mathsf{T}} $, the KKT condition can be expressed as,
\begin{align*}
	\nabla_\mathbf{w} \ell(x,\mathbf{w}) + 2 f(x,\mathbf{w})^{\mathsf{T}} \mathbf{P} f(x,\mathbf{w}) + \lambda^{\mathsf{T}} \nabla_{\mathbf{w}}g(x,\mathbf{w}) & =0 \\
	g(x,\mathbf{w}) &\leq 0\\
	 \lambda^{\mathsf{T}} g(x,\mathbf{w}) & = 0\\
	 \lambda &\geq 0
\end{align*}
Suppose the optimal state-action data pairs $ \mathcal{D}:=  \{(x_{i},\mathbf{w}_{i})\} := \{(x_{i},[u_{i},z_{i}]^{\mathsf{T}})\}$ obtained by querying the full MINMPC \eqref{Eq:MINMPC} satisfies the KKT conditions
of the myopic MPC stated above, then the myopic MPC policy is said to be consistent with the dataset $\mathcal{D}$.
In other words, we want to find $ \mathbf{P} $, such the the KKT conditions evaluated at the datapoints $ \{(x_{i},\mathbf{w}_{i})\} $ are close to zero.

To this end, the problem of imputing the cost-to-go function can be formulated as
\begin{subequations}\label{Eq:SDP}
	\begin{align}
		\min_{P,\{\lambda_{i}\}}& ~ \sum {}_{i=1}^M \|r_{\textrm{stat}}(x_{i},\mathbf{w}_{i})\|_2^2 + \|r_{\textrm{comp}}(x_{i},\mathbf{w}_{i})\|_2^2\label{eq:iocobjective} \\
		 \textrm{s.t.}  \; &  P\succeq 0, \\
		& \lambda_{i} \succeq 0, \; \; \forall i = 1,\dots,M \label{eq:possemidef}\\
	\end{align} \label{eq:IOC}
\end{subequations}
where KKT residuals for the $ i^{th} $ datapoint is  defined as
\begin{subequations}
\begin{align}
    r_{\textrm{stat}}(x_{i},\mathbf{w}_{i}) &=	\nabla_\mathbf{w} \ell(x_{i},\mathbf{w}_{i}) + 2 f(x_{i},\mathbf{w}_{i})^{\mathsf{T}} \mathbf{P} f(x_{i},\mathbf{w}_{i}) \nonumber \\
     & \qquad + \lambda_{i}^{\mathsf{T}} \nabla_{\mathbf{w}}g(x_{i},\mathbf{w}_{i})  \label{eq:aprxstatcond} \\
    r_{\textrm{comp}}(x_{i},\mathbf{w}_{i}) &=\, \lambda_{i}^{\mathsf{T}} g(x_{i},\mathbf{w}_{i}) \label{eq:aprxcompcond}
\end{align} \label{eq:aprxoptconds}
\end{subequations}
Note that offline problem of imputing the cost function \eqref{Eq:SDP} is a semi-definite program (SDP) which can be solved efficiently. The solution $P$ can then be used in a myopic MPC controller \eqref{Eq:ADPMINMPC1} with a prediction horizon as small as  $N=1$ that approximates the control decisions of the original long prediction horizon MINMPC controller \eqref{Eq:MINMPC}.

\section{Illustrative example}
\label{sec:benchmarkProblems}

To demonstrate the efficacy of myopic MPC with IOC-imputed cost-to-go weighting parameters, we apply this method to the Lokta-Volterra fishing problem, which \cite{sager2006} present as a foundational benchmark problem for mixed-integer control of hybrid nonlinear systems. This benchmark problem  seeks to bring the oscillations of predator and prey populations close to a steady-state solution through an optimal fishing allowance. The dynamic system \eqref{eq:fishingdynamics} describes the evolution of the prey population $x_1$ and the predator population $x_2$ as
\begin{subequations}
\begin{gather}
 \begin{bmatrix} \dot{x_1}  \\ \dot{x_2} \end{bmatrix}
 =
  \begin{bmatrix}
   x_1 - x_1x_2 - c_1x_1u \\
   -x_2 + x_1x_2 - c_2x_2u
   \end{bmatrix},
   \label{eq:fishingdynamics} \\
   u \in \{0,1\}, \label{eq:binary} \\
   [x_1, x_2]^\top \geq 0. \label{eq:fishcons}
\end{gather}
\label{eq:fishingproblem}
\end{subequations}
where the growth of both populations are coupled and partly controlled by fishing rates governed with coefficients $c_1$ and $c_2$. The system includes a binary input variable $z_k = u$ \eqref{eq:binary} representing the decision to fish ($u=1$) or not to fish ($u=0$) and a state constraint \eqref{eq:fishcons} requiring the fish populations to be positive.





 We first formulate this decision-making problem as a mixed-integer optimal control problem of the form\eqref{Eq:MINMPC}. The controller objective parameters are tuned to achieve good reference tracking for a prediction horizon length $N = ??$ that approximates the solution of $N=\infty$ with discrete sampling time $t_s$. 
 We construct the controller in \texttt{Matlab} using the \texttt{CasADi} toolbox \citep{Andersson2018CasADi} and the \texttt{Bonmin} mixed integer nonlinear programming solver \citep{bonami2008algorithmic}. 
 The resulting solution set size $M=120$ required $t_{\textrm{CPU}}=2.07\times10^3$ ({s}) to solve on an Intel(R) Core\textsuperscript{TM} i5 processor running at 2.40 GHz with 15.7 GB of useable RAM.

 Since this is not amenable for online implementation, we aim to impute a convex cost-to-go function $ \mathcal{V}(x):= x^{\mathsf{T}} \mathbf{P} x$, such that the we can solve a myopic MPC online with $ N = 1 $ and $ \mathcal{V}(x) $ as the cost-to-go function. To compute $ \mathbf{P} $, the MINMPC controller then is run in an offline feedback control simulation with perfect state observation,  for three initial states, resulting in 3 trajectories of data $\mathcal{D}$. 

%
%


\subsection{Imputing the cost-to-go offline}
\label{subsec:imputedParam}

Using the state-action data set $ \mathcal{D} $  obtained from the offline MINMPC simulations, we solve the SDP for $P$ that minimizes $r_{stat}^{(m)}$ and $r_{comp}^{(m)}$ \eqref{Eq:SDP}, yielding

\begin{gather*}
 P = \begin{bmatrix} 3.07\times10^{-4} & 2.13\times10^{-5} \\ 2.13\times10^{-5}  & 2.06\times10^{-3} \end{bmatrix},
   \label{eq:Pmatrix}
\end{gather*}

and $\|r_{stat}\|_\infty = 1.29\times10^{-6}$ and $\|r_{comp}\|_\infty = 1.91\times10^{-6}$. As required in \eqref{eq:IOC}, the $P$ matrix is positive definite. Moreover, the maximum residuals for both the stationarity and complementary slackness KKT conditions are vanishingly small. We conclude, therefore, that the imputed cost-to-go parameter $P$ is consistent with all observed solutions to the Lotka-Volterra fishing MINMPC problem. Consequently, the $P$ matrix is appropriate as a tuning matrix to build the myopic MPC controller in \eqref{Eq:ADPMINMPC1}.

\subsection{Online Controller Performance}
\label{subsec:refTracking}

\begin{figure}[t]
\centering
\parbox{\columnwidth}{\includegraphics[width=\columnwidth]{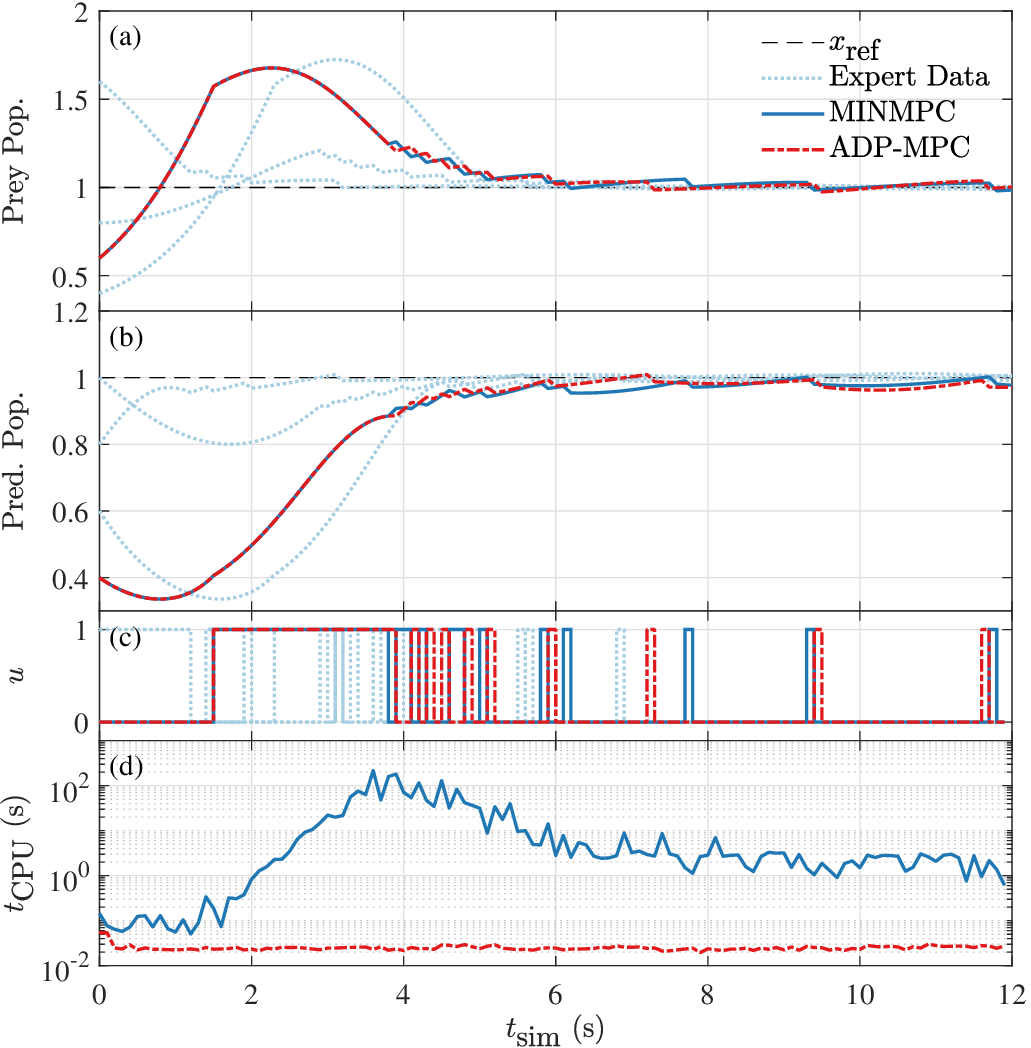}}
\caption{(a) Prey and (b) predator state population for the three training solution set trajectories (in light blue) computed offline (without plant-model mismatch and measurement noise), the trajectory computed with MINMPC controller (in blue), and the trajectory computed with the myopic MPC controller (in red). The reference is indicated by a black, dashed line. (c) The control decisions for both the training sets and the online controllers. (d) The computation time per controller decision.}
\label{fig:fishing}
\end{figure}

The goal for the MPC controller is to track a state reference without violating system constraints. Within the context of this work, however, we introduce an additional key performance goal of the maximum computation time per controller decision. The controller performance is tested in a discrete time simulation of the dynamic system, with added plant-model mismatch (implemented as $10\%$ error in the fishing coefficients $c_1$ and $c_2$ of \eqref{eq:fishingdynamics}) and synthetic measurement noise (implemented as a zero-mean Gaussian white noise). The results of the control simulations for the Lotka-Volterra fishing problem (including the offline expert demonstrations) are given in Fig. \ref{fig:fishing}.


As shown in Fig. \ref{fig:fishing}, the myopic MPC controller using a $1$ step prediction horizon and the cost-to-go parameter imputed using IOC closely reproduces the control actions computed by the full horizon MINMPC controller with slight deviation. It is worth noting here the MINMPC controller does not drive the state to the state reference unless $N\gg 1$, meaning that the imputed cost-to-go parameter is responsible for controller performance.

While the ability of the myopic MPC controller to achieve reference tracking performance that approximates that of the long prediction horizon controller is satisfactory, what is more striking is the reduction in computation time required to solve the control problem. Fig. \ref{fig:fishing}d shows the controller computation time at each simulation step. The MINMPC controller was very computationally expensive, requiring a maximum computation time of $t_{\textrm{CPU}}=217$ ({s}) per decision to solve. However, the maximum computation time of the myopic MPC controller was only $t_{\textrm{CPU}}= 54.1$ ({ms}). By greatly reducing the control problem complexity, the myopic MPC controller was able to reduce the maximum computation time per controller decision by more than three orders of magnitude over the MINMPC controller. Hence, the myopic MPC method takes a controller that is computationally intractable and makes it solvable in real time. The simulations were performed on an Intel(R) Core\textsuperscript{TM} i5 processor running at 2.40 GHz with 15.7 GB of useable RAM.

\section{Conclusion}
\label{sec:conclusion}

In this work, we have demonstrated that the KKT residual minimization method for IOC can impute the cost-to-go weighting parameter of a $1$-step prediction horizon MPC problem from a set of action-state pair solutions computed offline by a long prediction horizon MINMPC controller. This parameter is shown to be consistent with the offline solution set through the definition of approximate optimality. We then show that the myopic MPC controller using this cost-to-go parameter closely replicates the reference tracking performance of the original MPC controller. The method that we demonstrate in this paper offers a powerful tool for simplifying computationally complex MINMPC controllers.

There remain several open questions that this proof-of-concept demonstration raises. While we impute a parameter for a quadratic cost-to-go term because it leads to an IOC that is solvable as an SDP, a quadratic term may not be adequate approximate the cost-to-go of certain MPC problems. Bellman's principle of optimality indicates that some function must exist that is consistent with a given solution set of expert controller demonstrations. Broadening the definition of the imputed objective term to any convex function and solving the same IOC problem would broaden the applicability of the method described here. Additionally, complete characterization of stability and recursive feasibility properties of the myopic MPC controller is outside of the scope of this paper. Nonetheless, the authors consider such a characterization invaluable and plan to explore it with the future work.

Lastly, the question of computational costs of the offline solution set is not addressed in this work. While the cost of computing offline solutions was acceptable for the two simple benchmark problems shown here, this may not be the case for control problems with larger state spaces, not to mention hybrid systems with combinatorial integer actuators. The authors seek to combine the imputed objective method shown here with methods to make offline data generation more efficient. One example of this may be data augmentation method demonstrated by \cite{DK2021DataAug}. Answering the above open research questions would hopefully yield a generalized approach to reducing controller complexity with stability and feasibility guarantees from efficiently generated training data for any MPC problem.

\bibliography{BibL4DC}             

\begin{thebibliography}{10}
\providecommand{\natexlab}[1]{#1}
\providecommand{\url}[1]{\texttt{#1}}
\providecommand{\urlprefix}{URL }
\expandafter\ifx\csname urlstyle\endcsname\relax
  \providecommand{\doi}[1]{doi:\discretionary{}{}{}#1}\else
  \providecommand{\doi}{doi:\discretionary{}{}{}\begingroup
  \urlstyle{rm}\Url}\fi

\bibitem[{Andersson et~al.(In Press, 2018)Andersson, Gillis, Horn, Rawlings,
  and Diehl}]{Andersson2018CasADi}
Andersson, J.A.E., Gillis, J., Horn, G., Rawlings, J.B., and Diehl, M. (In
  Press, 2018).
\newblock {CasADi} -- {A} software framework for nonlinear optimization and
  optimal control.
\newblock \emph{Mathematical Programming Computation}.

\bibitem[{Bertsekas(2011)}]{bertsekas2011dynamic}
Bertsekas, D.P. (2011).
\newblock Dynamic programming and optimal control 3rd edition, volume ii.
\newblock \emph{Belmont, MA: Athena Scientific}.

\bibitem[{Bonami et~al.(2008)Bonami, Biegler, Conn, Cornu{\'e}jols, Grossmann,
  Laird, Lee, Lodi, Margot, Sawaya et~al.}]{bonami2008algorithmic}
Bonami, P., Biegler, L.T., Conn, A.R., Cornu{\'e}jols, G., Grossmann, I.E.,
  Laird, C.D., Lee, J., Lodi, A., Margot, F., Sawaya, N., et~al. (2008).
\newblock An algorithmic framework for convex mixed integer nonlinear programs.
\newblock \emph{Discrete optimization}, 5(2), 186--204.

\bibitem[{De~Mauri et~al.(2020)De~Mauri, Van~Roy, Gillis, Swevers, and
  Pipeleers}]{deMauri2020real}
De~Mauri, M., Van~Roy, W., Gillis, J., Swevers, J., and Pipeleers, G. (2020).
\newblock Real time iterations for mixed-integer model predictive control.
\newblock In \emph{2020 European Control Conference (ECC)}, 699--705. IEEE.

\bibitem[{Karg and Lucia(2018)}]{karg2018deep}
Karg, B. and Lucia, S. (2018).
\newblock Deep learning-based embedded mixed-integer model predictive control.
\newblock In \emph{2018 European Control Conference (ECC)}, 2075--2080. IEEE.

\bibitem[{Keshavarz et~al.(2011)Keshavarz, Wang, and
  Boyd}]{keshavarz2011imputing}
Keshavarz, A., Wang, Y., and Boyd, S. (2011).
\newblock Imputing a convex objective function.
\newblock In \emph{2011 IEEE international symposium on intelligent control},
  613--619. IEEE.

\bibitem[{Krishnamoorthy(2021)}]{DK2021DataAug}
Krishnamoorthy, D. (2021).
\newblock A sensitivity-based data augmentation framework for model predictive
  control policy approximation.
\newblock \emph{IEEE Transactions on Automatic Control}, {In-Press}.

\bibitem[{Richards and How(2005)}]{richards2005MIP-MPC}
Richards, A. and How, J. (2005).
\newblock Mixed-integer programming for control.
\newblock In \emph{American Control Conference, 2005. Proceedings of the 2005},
  2676--2683. IEEE.

\bibitem[{Sager(2009)}]{sager2009reformulations}
Sager, S. (2009).
\newblock Reformulations and algorithms for the optimization of switching
  decisions in nonlinear optimal control.
\newblock \emph{Journal of Process Control}, 19(8), 1238--1247.

\bibitem[{Sager et~al.(2006)Sager, Bock, Diehl, Reinelt, and
  Schloder}]{sager2006numerical}
Sager, S., Bock, H.G., Diehl, M., Reinelt, G., and Schloder, J.P. (2006).
\newblock Numerical methods for optimal control with binary control functions
  applied to a lotka-volterra type fishing problem.
\newblock In \emph{Recent Advances in Optimization}, 269--289. Springer.

\end{thebibliography}








\end{document}